\documentclass[runningheads]{llncs}
\usepackage{graphicx}
\usepackage{amssymb}
\usepackage{amsmath}
\usepackage{caption}
\usepackage{subcaption}
\usepackage{tabularx}
\usepackage{hyperref}
\usepackage[dvipsnames]{xcolor}

\newcommand{\norm}[1]{\left\lVert#1\right\rVert}
\newcolumntype{Y}{>{\centering\arraybackslash}X}

\DeclareMathOperator*{\argmin}{arg\,min}

\begin{document}
%
\title{Deep J-Sense: Accelerated MRI Reconstruction via Unrolled Alternating Optimization\thanks{Part of this work has been accepted for presentation at the ISMRM Annual Conference 2021. Supported by ONR grant N00014-19-1-2590, NIH Grant U24EB029240, and an AWS Machine Learning Research Award.}}


\titlerunning{Deep J-Sense}
%
\author{Marius Arvinte\inst{1} \and
Sriram Vishwanath\inst{1} \and
Ahmed H. Tewfik\inst{1} \and
Jonathan I. Tamir\inst{1}}

\authorrunning{M. Arvinte et al.}

\institute{The University of Texas at Austin, Austin TX 78705, USA}

\maketitle              

\begin{abstract}
Accelerated multi-coil magnetic resonance imaging reconstruction has seen a substantial recent improvement combining compressed sensing with deep learning. However, most of these methods rely on estimates of the coil sensitivity profiles, or on calibration data for estimating model parameters. Prior work has shown that these methods degrade in performance when the quality of these estimators are poor or when the scan parameters differ from the training conditions. Here we introduce Deep J-Sense as a deep learning approach that builds on unrolled alternating minimization and increases robustness: our algorithm refines both the magnetization (image) kernel and the coil sensitivity maps. Experimental results on a subset of the knee fastMRI dataset show that this increases reconstruction performance and provides a significant degree of robustness to varying acceleration factors and calibration region sizes.

\end{abstract}

\section{Introduction}
Parallel MRI is a multi-coil acceleration technique that is standard in nearly all clinical systems \cite{pruessmann1999sense,griswold2002generalized,sodicksonssmash}. The technique uses multiple receive coils to measure the signal in parallel, and thus accelerate the overall acquisition. Compressed sensing-based methods with suitably chosen priors have constituted one of the main drivers of progress in parallel MRI reconstruction for the past two decades \cite{lustig2007sparse,deshmane2012parallel,uecker2014espirit,rosenzweig2018simultaneous}. While parallel MRI provides additional degrees of freedom via simultaneous measurements, it brings its own set of challenges related to estimating the spatially varying \textit{sensitivity maps} of the coils, either explicitly \cite{ying2007joint,uecker2014espirit} or implicitly \cite{griswold2002generalized}. These algorithms typically use a fully sampled region of k-space or a low-resolution reference scan as an auto-calibration signal (ACS), either to estimate k-space kernels \cite{griswold2002generalized,lustig2010spirit}, or to estimate coil sensitivity profiles \cite{pruessmann1999sense}. Calibration-free methods have been proposed that leverage structure in the parallel MRI model; namely, that sensitivity maps smoothly vary in space \cite{ying2007joint,rosenzweig2018simultaneous} and impose low-rank structure \cite{haldar2013low,shin2014calibrationless}.

Deep learning has recently enabled significant improvement to image quality for accelerated MRI when combined with ideas from compressed sensing in the form of \textit{unrolled iterative optimization} \cite{schlemper2017deep,hammernik2018learning,aggarwal2018modl,sriram2020end,Sriram_2020_CVPR}. Our work falls in this category, where learnable models are interleaved with optimization steps and the entire system is trained end-to-end with a supervised loss. However, there are still major open questions concerning the robustness of these models, especially when faced with distributional shifts \cite{antun2020instabilities}, i.e., when the scan parameters at test time do not match the ones at training time or the robustness of methods across different training conditions. This is especially prudent for models that use estimated sensitivity maps, and thus require reliable estimates as input.

Our contributions are the following: i) we introduce a novel deep learning-based parallel MRI (pMRI) reconstruction algorithm that unrolls an alternating optimization to jointly solve for the image and sensitivity map kernels directly in k-space; ii) we train and evaluate our model on a subset of the fastMRI knee dataset and show improvements in reconstruction fidelity; and iii) we evaluate the robustness of our proposed method on distributional shifts produced by different sampling parameters and obtain state-of-the-art performance. An open-source implementation of our method is publicly available\footnote{\href{https://github.com/utcsilab/deep-jsense}{\texttt{https://github.com/utcsilab/deep-jsense}}}.

\section{System Model and Related Work}
In parallel MRI, the signal is measured by an array of radio-frequency receive coils distributed around the body, each with a spatially-varying sensitivity profile. In the measurement model, the image is linearly mixed with each coil sensitivity profile and sampled in the Fourier domain (k-space). Scans can be accelerated by reducing the number of acquired k-space measurements, and solving the inverse problem by leveraging redundancy across the receive channels as well as redundancy in the image representation.

We consider pMRI acquisition with $C$ coils. Let $\mathbf{m} \in \mathbb{C}^{n}$ and $\mathbf{s} \in \mathbb{C}^{C\times k}$ be the $n$-dimensional image (magnetization) and set of $k$-dimensional sensitivity map kernels, respectively, defined directly in k-space. We assume that $\mathbf{k}_i$, the k-space data of the $i$-th coil image, is given by the linear convolution between the two kernels as
\begin{equation}
\mathbf{k}_i = \mathbf{s}_i * \mathbf{m}.
\label{eq:fw_model}
\end{equation}

Joint image and map reconstruction formulates \eqref{eq:fw_model} as an optimization problem where both variables are unknown. Given a sampling mask represented by the matrix $\mathbf{A}$ and letting $\mathbf{y} = \mathbf{A}\mathbf{k} + \mathbf{n}$ be the sampled noisy multicoil k-space data, where $\mathbf{k}$ is the ground-truth k-space data and $\mathbf{n}$ is the additive noise, the optimization problem is
\begin{equation}
\argmin_{\mathbf{s}, \mathbf{m}} \frac{1}{2}\norm{\mathbf{y} - \mathbf{A} (\mathbf{s} * \mathbf{m})}_2^2 + \lambda_{\mathbf{m}} R_{\mathbf{m}}(\mathbf{m}) + \lambda_{\mathbf{s}} R_{\mathbf{s}}(\mathbf{s}).
\label{eq:two_vars}
\end{equation}

$R_{\mathbf{m}}(\mathbf{m})$ and $R_{\mathbf{s}}(\mathbf{s})$ are regularization terms that enforce priors on the two variables, e.g., J-Sense \cite{ying2007joint} uses polynomial regularization for $\mathbf{s}$. We let $\mathbf{A}_{\mathbf{m}}$ and $\mathbf{A}_{\mathbf{s}}$ denote the linear operators composed by convolution with the fixed variable and $\mathbf{A}$. The generic solution of \eqref{eq:two_vars} by alternating minimization involves the steps
\begin{subequations}
\begin{align}
\mathbf{m} & = \argmin_{\mathbf{m}} \frac{1}{2}\norm{\mathbf{y} - \mathbf{A}_\mathbf{m} \mathbf{m}}^2_2 + \lambda_{\mathbf{m}} R_{\mathbf{m}}(\mathbf{m}), \\
\mathbf{s} & = \argmin_{\mathbf{s}} \frac{1}{2}\norm{\mathbf{y} - \mathbf{A}_\mathbf{s} \mathbf{s}}_2^2 + \lambda_{\mathbf{s}} R_{\mathbf{s}}(\mathbf{s}).
\end{align}
\end{subequations}
If L2-regularization is used for the $R$ terms, each sub-problem is a linear least squares minimization, and the Conjugate Gradient (CG) algorithm \cite{shewchuk1994introduction} with a fixed number of steps can be applied to obtain an approximate solution.

\subsection{Deep Learning for MRI Reconstruction}
Model-based deep learning architectures for accelerated MRI reconstruction have recently demonstrated state-of-the-art performance in \cite{hammernik2018learning,aggarwal2018modl,sriram2020end}. The MoDL algorithm \cite{aggarwal2018modl} in particular is used to solve \eqref{eq:fw_model} when only the image kernel variable is unknown and a deep neural network $\mathcal{D}$ is used in $R_{\mathbf{m}}(\mathbf{m})$. The formulation of the optimization problem is 
\begin{equation}
\argmin_{\mathbf{m}} \frac{1}{2}\norm{\mathbf{y} - \mathbf{A}_{\mathbf{m}} \mathbf{m}}_2^2 + \lambda \norm{\mathcal{D}(\mathbf{m}) - \mathbf{m}}_2^2.
\label{eq:modl}
\end{equation}

To unroll the optimization in \eqref{eq:modl}, the authors split each step in two different sub-problems. The first sub-problem treats $\mathcal{D}(\mathbf{m})$ as a constant and uses the CG algorithm to update $\mathbf{m}$. The second sub-problem treats $\mathcal{D}$ as a proximal operator and is solved by direct assignment, i.e., $\mathbf{m}^{+} = \mathcal{D}(\mathbf{m})$. In our work, we use the same approach for unrolling the optimization, but we use a pair of deep neural networks, one for each variable in \eqref{eq:two_vars}. Unlike \cite{aggarwal2018modl}, our work does not rely on a pre-computed estimate of the sensitivity maps, but instead treats them as an optimization variable.

The idea of learning a sensitivity map estimator was first described in \cite{cheng2019ismrm}. Recently, the work in \cite{sriram2020end} introduces the E2E-VarNet architecture that addresses the issue of estimating the sensitivity maps by training a \textit{sensitivity map estimation} module in the form of a deep neural network. Like the ESPiRiT algorithm \cite{uecker2014espirit}, E2E-VarNet additionally enforces that the sensitivity maps are normalized per-pixel. The sensitivity maps are held fixed during the unroll. 

The architecture -- which uses gradient descent instead of CG -- is then trained end-to-end, using the estimated sensitivity maps and the forward operator $\mathbf{A}_\mathbf{m}$. The major difference between our work and \cite{sriram2020end} is that we iteratively update the maps instead of using a single-shot data-based approach \cite{sriram2020end} or the ESPiRiT algorithm \cite{hammernik2018learning,aggarwal2018modl}. As our results show in the sequel, this has a significant impact on the out-of-distribution robustness of the approach on scans whose parameters differ from the training set. Finally, there is a recent mention of jointly optimizing the image and sensitivity maps in the IC-Net architecture \cite{muckley2020state}, but no associated paper or implementation details are currently available.

\section{Deep J-Sense: Unrolled Alternating Optimization}
We unroll the optimization in \eqref{eq:two_vars} by alternating between optimizing the two variables as
\begin{subequations}
\begin{align}
\mathbf{s}^{} & = \argmin_{\mathbf{s}} \frac{1}{2}\norm{\mathbf{y} - \mathbf{A}_\mathbf{s} \mathbf{s}}_2^2
+ \lambda_\mathbf{s} R_{\mathbf{s}}(\mathbf{s}),\\
\mathbf{s}^{+} & = \mathcal{D}_\mathbf{s}^{} (\mathbf{s}^{}),\\
\mathbf{m}^{} & = \argmin_{\mathbf{m}} \frac{1}{2}\norm{\mathbf{y} - \mathbf{A}_\mathbf{m} \mathbf{m}}_2^2
+ \lambda_\mathbf{m} R_{\mathbf{m}}(\mathbf{m}),\\
\mathbf{m}^{+} & = \mathcal{D}_\mathbf{m}^{} (\mathbf{m}^{}),
\end{align}
\label{eq:our_unrolls}
\end{subequations}

\noindent where $R$ is defined as
\begin{equation}
    R(\mathbf{x}) = \norm{\mathcal{F}\{\mathcal{D}(\mathcal{F}^{-1}\{ \mathbf{x} \} )\} - \mathbf{x}}_2^2,
\end{equation}

\noindent for both $\mathbf{m}$ and $\mathbf{s}$, $\mathcal{F}$ is the Fourier transform, and $\mathcal{D}$ is the deep neural network corresponding to each variable. Similar to MoDL, we set $\mathcal D^{(j)}_{\mathbf m} = \mathcal{D}_\mathbf{m}$ and $\mathcal D^{(j)}_{\mathbf s} = \mathcal{D}_\mathbf{s}$ across all unrolls, leading to the efficient use of learnable weights. The coefficients $\lambda_\mathbf{s}$ and $\lambda_\mathbf{m}$ are also learnable.
\begin{figure}
\centering
\includegraphics[width=0.95\textwidth]{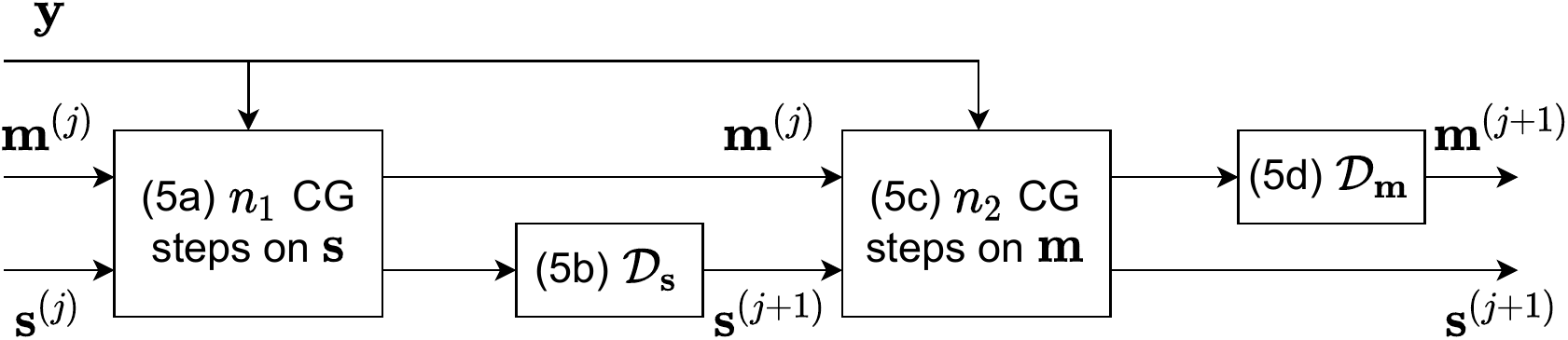}
\caption{A single unroll of the proposed scheme. The CG algorithm is executed on the loss given by the undersampled data $\mathbf{y}$ and the measurement matrices $\mathbf{A}_\mathbf{s}$ and $\mathbf{A}_\mathbf{m}$. Each block is matched to its corresponding equation.}
\label{fig:block_diagram}
\end{figure}

The optimization is initialized with $\mathbf{m}^{(0)}$ and $\mathbf{s}^{(0)}$, obtained using a simple root sum-of-squares (RSS) estimate. Steps (5a) and (5c) are approximately solved with $n_1$ and $n_2$ steps of the CG algorithm, respectively, while steps (5b) and (5d) represent direct assignments. The two neural networks serve as generalized denoisers applied in the image domain and are trained in an end-to-end fashion after unrolling the alternating optimization for a number of $N$ outer steps. A block diagram of one unroll is shown in Fig. \ref{fig:block_diagram}.

Using the estimated image and map kernels after $N$ outer steps, we train the end-to-end network using the estimated RSS image as
\begin{equation}
    \hat{\mathbf{x}} = \sqrt{\sum_{i=1}^{C} | \mathcal{F}^{-1} \{ \mathbf{s}^{(N)}_i * \mathbf{m}^{(N)} \} | ^ 2},
\end{equation}

\noindent where the supervised loss is the structural similarity index (SSIM) between the estimated image and ground truth image as $L = -\text{SSIM}(\hat{x}, x)$.

Our model is implemented as a linear convolution of the image and maps kernels and can be seen as a unification of MoDL and J-Sense. For example, by setting $n_1 = 0$ and $\mathcal{D}_{\mathbf{s}}(\mathbf{s}) = \mathbf{s}$, the sensitivity maps are never updated and the proposed approach becomes MoDL. At the same time, removing the deep neural networks by setting $\mathcal{D}(\cdot) = 0$ and removing steps (5b) and (5d) leads to L2-regularized J-Sense. One important difference is that, unlike the ESPiRiT algorithm or E2E-VarNet, our model does \textit{not} perform pixel-wise normalization of the sensitivity maps, and thus does not impart a spatially-varying weighting across the image. Furthermore, since our forward model is in k-space, we use a small-sized kernel for the sensitivity maps as an implicit smoothness regularizer and reduction in memory. 

\section{Experimental Results}
We compare the performance of the proposed approach against MoDL \cite{aggarwal2018modl} and E2E-VarNet \cite{sriram2020end}. We train and evaluate all methods on a subset of the fastMRI knee dataset \cite{zbontar2018fastMRI} to achieve reasonable computation times. For training, we use the five central slices from each scan in the training set, for a total of $4580$ training slices. For evaluation, we use the five central slices from each scan in the validation set, for a total of $950$ validation slices. All algorithms are implemented in PyTorch \cite{NEURIPS2019_9015} and SigPy \cite{sigpy}. Detailed architectural and hyper-parameter choices are given in the supplementary material.

To evaluate the impact of optimizing the sensitivity map kernel, we compare the performance of the proposed approach with a version of MoDL trained on the same data, and that uses the same number of unrolls (both inner and outer) and the same architecture for the image denoising network $\mathcal{D}_\mathbf{m}$. We compare our robust performance with E2E-VarNet trained on the same data, and having four times more parameters, to compensate for the run-time cost of updating the sensitivity maps.

\subsection{Performance on Matching Test-Time Conditions}
We compare the performance of our method with that of MoDL and E2E-VarNet when the test-time conditions match those at training time on knee data accelerated by a factor of $R = 4$. For MoDL, we use a denoising network with the same number of parameters as our image denoising network and the same number of outer and inner CG steps. We use sensitivity maps estimated by the ESPiRiT algorithm via the BART toolbox \cite{tamir2016generalized}, where a SURE-calibrated version \cite{iyer2020sure} is used to select the first threshold, and we set the eigenvalue threshold to zero so as to not unfairly penalize MoDL, since both evaluation metrics are influenced by background noise. For E2E-VarNet, we use the same number of $N = 6$ unrolls (called \textit{cascades} in \cite{sriram2020end}) and U-Nets for all refinement modules.

\begin{table}
\caption{Validation performance on a subset of the fastMRI knee dataset. Higher average/median SSIM (lower NMSE) indicates better performance. Lower standard deviations are an additional desired quality.}
\centering
\label{table:plain_performance}
\begin{tabularx}{\textwidth}{|Y||Y|Y|Y||Y|Y|Y|}
    \hline
    & Avg. & Med. & $\sigma$ & Avg. & Med. & $\sigma$ \\
    & SSIM & SSIM & SSIM & NMSE & NMSE & NMSE \\
    \hline
    \textbf{MoDL} & $0.814$ & $0.840$ & $0.115$ & $0.0164$ & $0.0087$ & $0.0724$ \\
    \hline
    \textbf{E2E} & $0.824$ & $0.851$ & $0.107$ & $0.0111$ & $0.0068$ & $0.0299$ \\
    \hline
    \textbf{Ours} & $\mathbf{0.832}$ & $\mathbf{0.857}$ & $\mathbf{0.104}$ & $\mathbf{0.0091}$ & $\mathbf{0.0064}$ & $\mathbf{0.0095}$ \\
    \hline
\end{tabularx}
\end{table}


\begin{figure}
\centering
\includegraphics[width=1\textwidth]{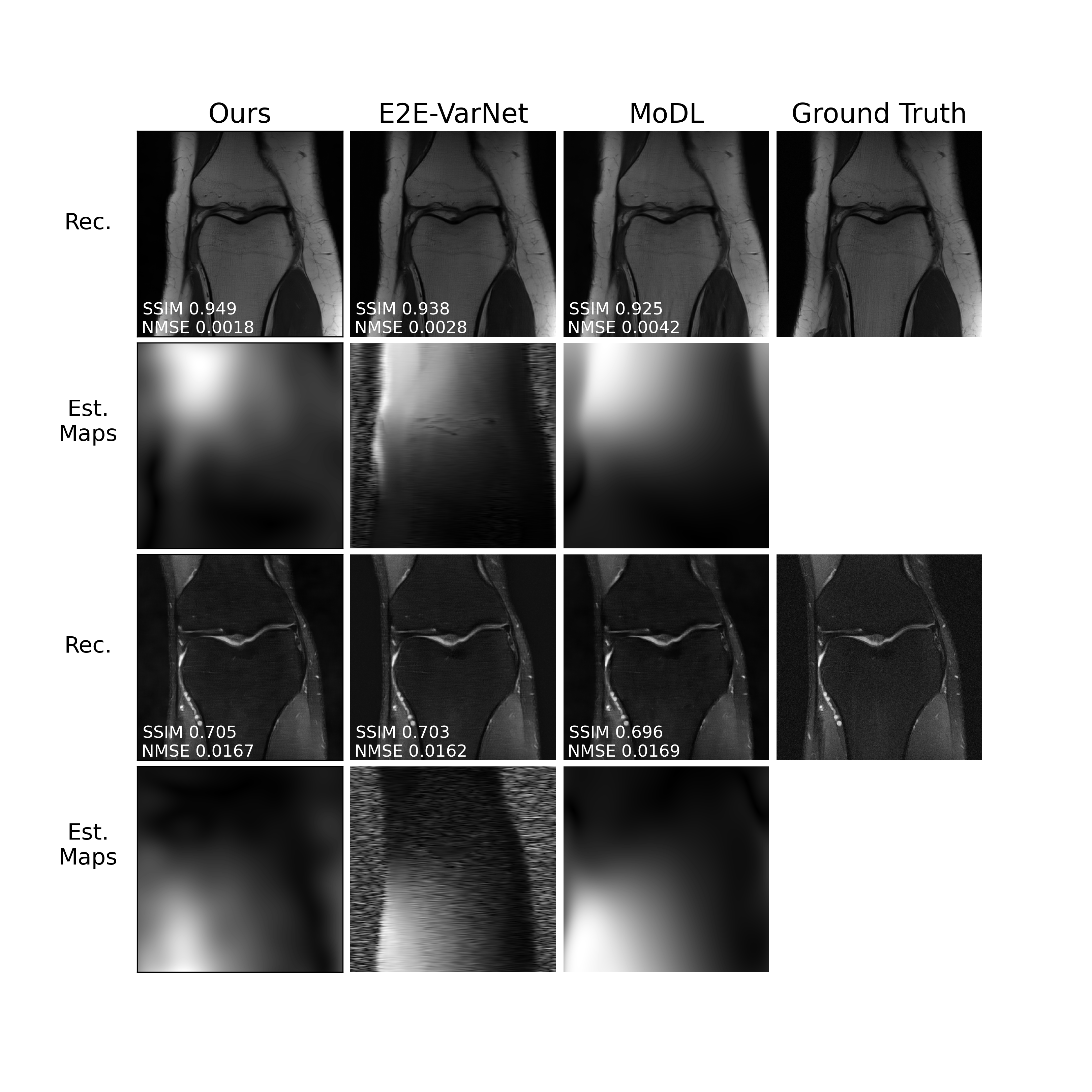}
\caption{Example reconstructions at $R=4$ and matching train-test conditions. The first and third rows represent RSS images from two different scans. The second and fourth rows represents the magnitude of the estimated sensitivity map (no ground truth available) for a random coil, from each scan, respectively. The maps under the MoDL column are estimated with the ESPiRiT algorithm.}
\label{fig:example_scan}
\end{figure}

Table \ref{table:plain_performance} shows statistical performance on the validation data. The comparison with MoDL allows us to evaluate the benefit of iteratively updating the sensitivity maps, which leads to a significant gain in both metrics. Furthermore, our method obtains a superior performance to E2E-VarNet while using \textit{four} times fewer trainable weights and the same number of outer unrolls. This demonstrates the benefit of trading off the number of parameters for computational complexity, since our model executes more CG iterations than both baselines. Importantly, Deep J-Sense shows a much lower variance of the reconstruction performance across the validation dataset, with nearly one order of magnitude gain against MoDL. After a detailed inspection, this is a direct effect of an increased performance gain on samples where MoDL fails at reconstruction, thus a strong argument for robustness.

Randomly chosen reconstructions of scans and the estimated sensitivity maps are shown in Fig. \ref{fig:example_scan}. We notice that our maps capture higher frequency components than those of MoDL (estimated via ESPiRiT), but do not contain spurious noise patterns outside the region spanned by the physical knee. In contrast, the maps from E2E-VarNet exhibit such patterns and, in the case of the second row, produce spurious patterns even \textit{inside} the region of interest, suggesting that the knee anatomy "leaks" in the sensitivity maps.

\subsection{Robustness to Varying Acceleration Factors}
\begin{figure}
\centering
\includegraphics[width=0.95\textwidth]{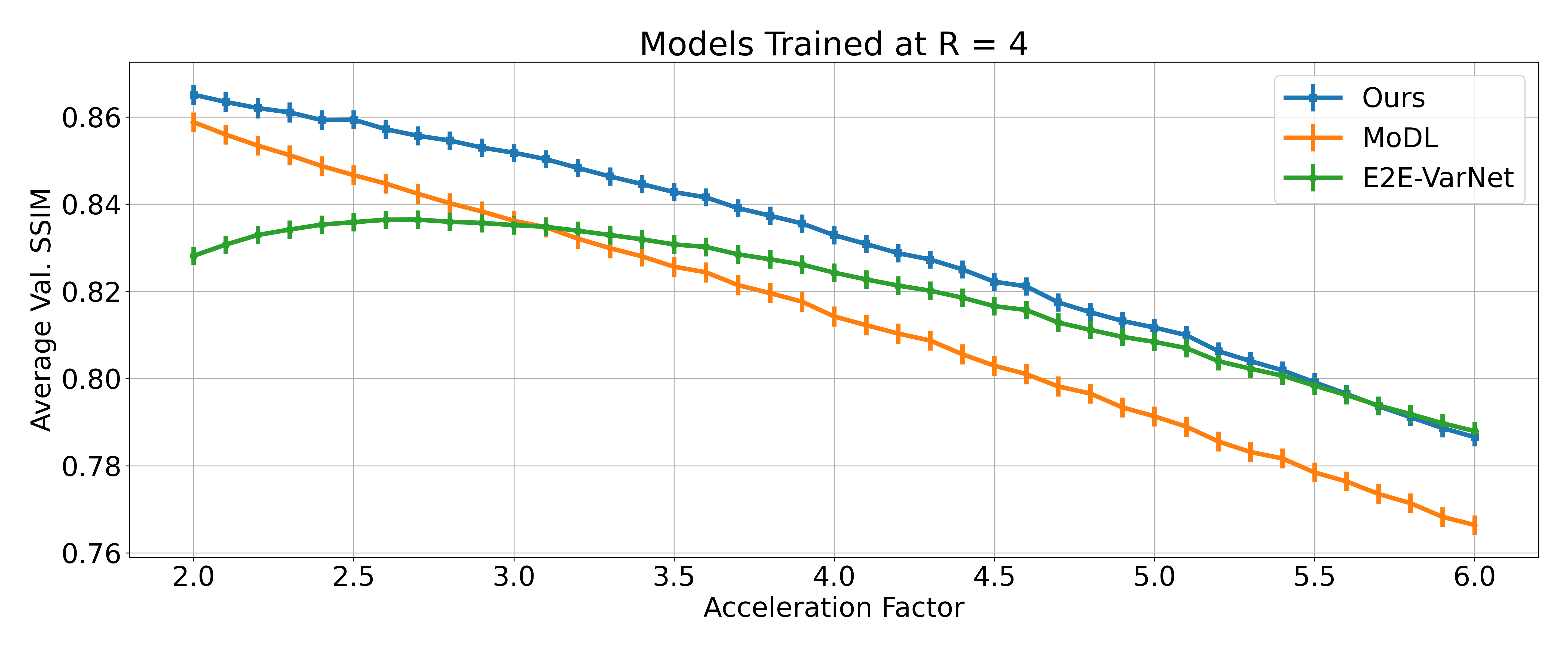}
\caption{Average SSIM on the fastMRI knee validation dataset evaluated at acceleration factors $R$ between $2$ and $6$ (with granularity $0.1$) using models trained at $R=4$. The vertical lines are proportional to the SSIM standard deviation in each case, from which no noticeable difference can be seen.}
\label{fig:robust}
\end{figure}

Fig. \ref{fig:robust} shows the performance obtained at acceleration factors between $2$ and $6$ All models are trained only on data with $R=4$, and the calibration region is kept at the same size for all tested factors. The modest performance gain for E2E-VarNet at $R<4$ confirms the findings in \cite{antun2020instabilities}, in that some models cannot efficiently use additional measurements if there is a train-test mismatch. At the same time, MoDL and the proposed method are able to overcome this effect, with our method significantly outperforming the baselines across all accelerations. Importantly, there is a significant decrease of the performance loss \textit{slope} against MoDL, rather than an additive gain, showing the benefit of estimating sensitivity maps using all the acquired measurements.

\subsection{Robustness to Varying ACS Size}
\begin{figure}
\centering
\includegraphics[width=0.95\textwidth]{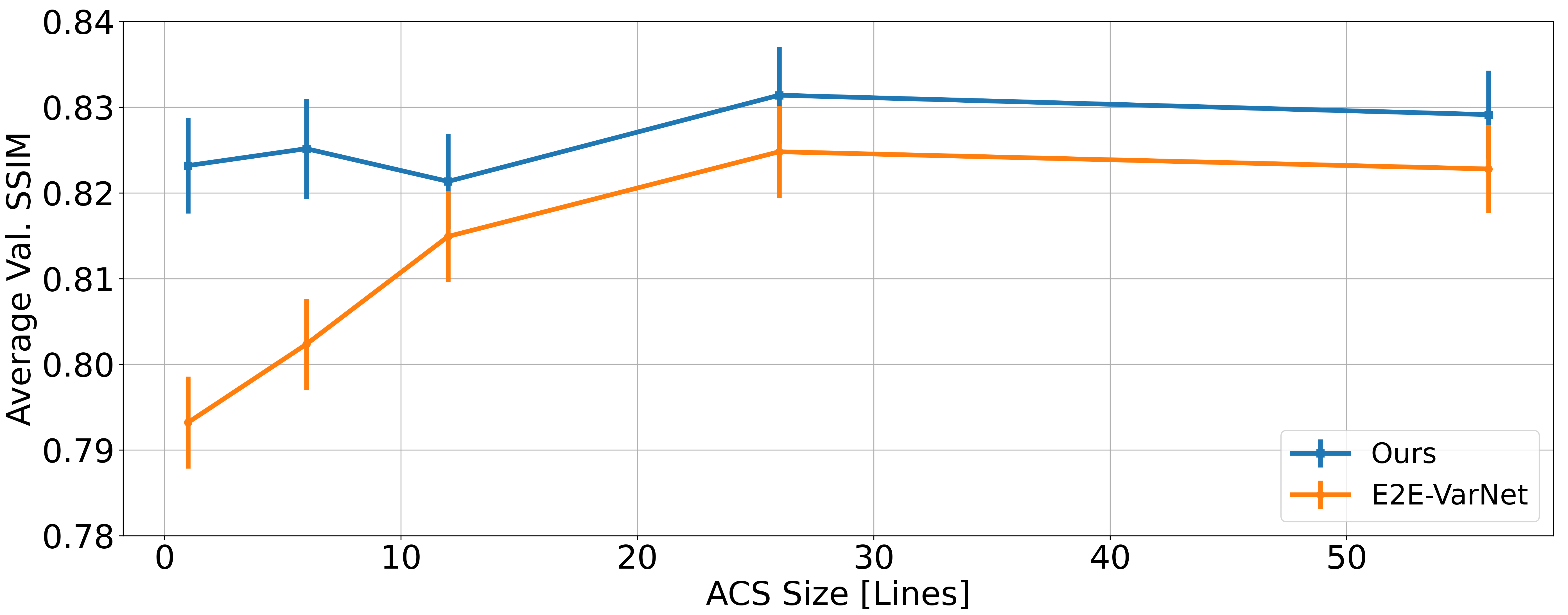}
\caption{Average SSIM on the fastMRI knee validation dataset evaluated at different sizes of the fully sampled auto-calibration region, at acceleration factor $R=4$. The vertical lines are proportional to the SSIM standard deviation in each case. Each model is trained and tested on the ACS size indicated on the x-axis.}
\label{fig:acs}
\end{figure}
We investigate the performance of the proposed method and E2E-VarNet as a function of the ACS region size (expressed as number of acquired lines in the phase encode direction). All models are trained at $R=4$ and ACS sizes $\{1, 6, 12, 26, 56\}$ and tested in the same conditions, giving a total of ten models. While there is no train-test mismatch in this experiment, Fig. \ref{fig:acs} shows a significant performance gain of the proposed approach when the calibration region is small (below six lines) and shows that overall, end-to-end performance is robust to the ACS size, with the drops at $1$ and $12$ lines not being statistically significant. This is in contrast to E2E-VarNet, which explicitly uses the ACS in its map estimation module and suffers a loss when this region is small.

\section{Discussion and Conclusions}
In this paper, we have introduced an end-to-end unrolled alternating optimization approach for accelerated parallel MRI reconstruction. Our approach jointly solves for the image and sensitivity map kernels directly in the k-space domain and generalizes several prior CS and deep learning methods. The results show that the method has superior reconstruction performance on a subset of the fastMRI knee dataset, and is robust to distributional shifts caused by varying acceleration factors and calibration region sizes. A possible extension of our work could include unrolling the estimates of multiple sets of sensitivity maps to account for scenarios with a reduced field of view \cite{uecker2014espirit,sandino2021accelerating}.

\newpage

\bibliographystyle{splncs04}
\bibliography{biblio}

\begin{thebibliography}{10}
\providecommand{\url}[1]{\texttt{#1}}
\providecommand{\urlprefix}{URL }
\providecommand{\doi}[1]{https://doi.org/#1}

\bibitem{aggarwal2018modl}
Aggarwal, H.K., Mani, M.P., Jacob, M.: Modl: Model-based deep learning
  architecture for inverse problems. IEEE transactions on medical imaging
  \textbf{38}(2),  394--405 (2018)

\bibitem{antun2020instabilities}
Antun, V., Renna, F., Poon, C., Adcock, B., Hansen, A.C.: On instabilities of
  deep learning in image reconstruction and the potential costs of ai.
  Proceedings of the National Academy of Sciences  \textbf{117}(48),
  30088--30095 (2020)

\bibitem{cheng2019ismrm}
Cheng, J.Y., Pauly, J.M., Vasanawala, S.S.: Multi-channel image reconstruction
  with latent coils and adversarial loss. ISMRM  (2019)

\bibitem{deshmane2012parallel}
Deshmane, A., Gulani, V., Griswold, M.A., Seiberlich, N.: Parallel mr imaging.
  Journal of Magnetic Resonance Imaging  \textbf{36}(1),  55--72 (2012)

\bibitem{griswold2002generalized}
Griswold, M.A., Jakob, P.M., Heidemann, R.M., Nittka, M., Jellus, V., Wang, J.,
  Kiefer, B., Haase, A.: Generalized autocalibrating partially parallel
  acquisitions (grappa). Magnetic Resonance in Medicine: An Official Journal of
  the International Society for Magnetic Resonance in Medicine  \textbf{47}(6),
   1202--1210 (2002)

\bibitem{haldar2013low}
Haldar, J.P.: Low-rank modeling of local $ k $-space neighborhoods (loraks) for
  constrained mri. IEEE transactions on medical imaging  \textbf{33}(3),
  668--681 (2013)

\bibitem{hammernik2018learning}
Hammernik, K., Klatzer, T., Kobler, E., Recht, M.P., Sodickson, D.K., Pock, T.,
  Knoll, F.: Learning a variational network for reconstruction of accelerated
  mri data. Magnetic resonance in medicine  \textbf{79}(6),  3055--3071 (2018)

\bibitem{iyer2020sure}
Iyer, S., Ong, F., Setsompop, K., Doneva, M., Lustig, M.: Sure-based automatic
  parameter selection for espirit calibration. Magnetic Resonance in Medicine
  \textbf{84}(6),  3423--3437 (2020)

\bibitem{lustig2007sparse}
Lustig, M., Donoho, D., Pauly, J.M.: Sparse mri: The application of compressed
  sensing for rapid mr imaging. Magnetic Resonance in Medicine: An Official
  Journal of the International Society for Magnetic Resonance in Medicine
  \textbf{58}(6),  1182--1195 (2007)

\bibitem{lustig2010spirit}
Lustig, M., Pauly, J.M.: Spirit: iterative self-consistent parallel imaging
  reconstruction from arbitrary k-space. Magnetic resonance in medicine
  \textbf{64}(2),  457--471 (2010)

\bibitem{muckley2020state}
Muckley, M.J., Riemenschneider, B., Radmanesh, A., Kim, S., Jeong, G., Ko, J.,
  Jun, Y., Shin, H., Hwang, D., Mostapha, M., et~al.: State-of-the-art machine
  learning mri reconstruction in 2020: Results of the second fastmri challenge.
  arXiv preprint arXiv:2012.06318  (2020)

\bibitem{sigpy}
Ong, F., Lustig, M.: Sigpy: A python package for high performance iterative
  reconstruction. Proceedings of the ISMRM 27th Annual Meeting, Montreal,
  Quebec, Canada pp. 4819--4819 (2019)

\bibitem{NEURIPS2019_9015}
Paszke, A., Gross, S., Massa, F., Lerer, A., Bradbury, J., Chanan, G., Killeen,
  T., Lin, Z., Gimelshein, N., Antiga, L., Desmaison, A., Kopf, A., Yang, E.,
  DeVito, Z., Raison, M., Tejani, A., Chilamkurthy, S., Steiner, B., Fang, L.,
  Bai, J., Chintala, S.: Pytorch: An imperative style, high-performance deep
  learning library. In: Wallach, H., Larochelle, H., Beygelzimer, A.,
  d\textquotesingle Alch\'{e}-Buc, F., Fox, E., Garnett, R. (eds.) Advances in
  Neural Information Processing Systems 32, pp. 8024--8035. Curran Associates,
  Inc. (2019),
  \url{http://papers.neurips.cc/paper/9015-pytorch-an-imperative-style-high-performance-deep-learning-library.pdf}

\bibitem{pruessmann1999sense}
Pruessmann, K.P., Weiger, M., Scheidegger, M.B., Boesiger, P.: Sense:
  sensitivity encoding for fast mri. Magnetic Resonance in Medicine: An
  Official Journal of the International Society for Magnetic Resonance in
  Medicine  \textbf{42}(5),  952--962 (1999)

\bibitem{rosenzweig2018simultaneous}
Rosenzweig, S., Holme, H.C.M., Wilke, R.N., Voit, D., Frahm, J., Uecker, M.:
  Simultaneous multi-slice mri using cartesian and radial flash and regularized
  nonlinear inversion: Sms-nlinv. Magnetic resonance in medicine
  \textbf{79}(4),  2057--2066 (2018)

\bibitem{sandino2021accelerating}
Sandino, C.M., Lai, P., Vasanawala, S.S., Cheng, J.Y.: Accelerating cardiac
  cine mri using a deep learning-based espirit reconstruction. Magnetic
  Resonance in Medicine  \textbf{85}(1),  152--167 (2021)

\bibitem{schlemper2017deep}
Schlemper, J., Caballero, J., Hajnal, J.V., Price, A.N., Rueckert, D.: A deep
  cascade of convolutional neural networks for dynamic mr image reconstruction.
  IEEE transactions on Medical Imaging  \textbf{37}(2),  491--503 (2017)

\bibitem{shewchuk1994introduction}
Shewchuk, J.R., et~al.: An introduction to the conjugate gradient method
  without the agonizing pain (1994)

\bibitem{shin2014calibrationless}
Shin, P.J., Larson, P.E., Ohliger, M.A., Elad, M., Pauly, J.M., Vigneron, D.B.,
  Lustig, M.: Calibrationless parallel imaging reconstruction based on
  structured low-rank matrix completion. Magnetic resonance in medicine
  \textbf{72}(4),  959--970 (2014)

\bibitem{sodicksonssmash}
Sodickson, D.K., Manning, W.J.: Simultaneous acquisition of spatial harmonics
  (smash): fast imaging with radiofrequency coil arrays. Magnetic resonance in
  medicine  \textbf{38}(4),  591--603 (1997)

\bibitem{sriram2020end}
Sriram, A., Zbontar, J., Murrell, T., Defazio, A., Zitnick, C.L., Yakubova, N.,
  Knoll, F., Johnson, P.: End-to-end variational networks for accelerated mri
  reconstruction. International Conference on Medical Image Computing and
  Computer-Assisted Intervention pp. 64--73 (2020)

\bibitem{Sriram_2020_CVPR}
Sriram, A., Zbontar, J., Murrell, T., Zitnick, C.L., Defazio, A., Sodickson,
  D.K.: Grappanet: Combining parallel imaging with deep learning for multi-coil
  mri reconstruction. Proceedings of the IEEE/CVF Conference on Computer Vision
  and Pattern Recognition (CVPR)  (June 2020)

\bibitem{tamir2016generalized}
Tamir, J.I., Ong, F., Cheng, J.Y., Uecker, M., Lustig, M.: Generalized magnetic
  resonance image reconstruction using the berkeley advanced reconstruction
  toolbox. In: ISMRM Workshop on Data Sampling and Image Reconstruction,
  Sedona, AZ (2016)

\bibitem{uecker2014espirit}
Uecker, M., Lai, P., Murphy, M.J., Virtue, P., Elad, M., Pauly, J.M.,
  Vasanawala, S.S., Lustig, M.: Espirit—an eigenvalue approach to
  autocalibrating parallel mri: where sense meets grappa. Magnetic resonance in
  medicine  \textbf{71}(3),  990--1001 (2014)

\bibitem{ying2007joint}
Ying, L., Sheng, J.: Joint image reconstruction and sensitivity estimation in
  sense (jsense). Magnetic Resonance in Medicine: An Official Journal of the
  International Society for Magnetic Resonance in Medicine  \textbf{57}(6),
  1196--1202 (2007)

\bibitem{zbontar2018fastMRI}
Zbontar, J., Knoll, F., Sriram, A., Murrell, T., Huang, Z., Muckley, M.J.,
  Defazio, A., Stern, R., Johnson, P., Bruno, M., Parente, M., Geras, K.J.,
  Katsnelson, J., Chandarana, H., Zhang, Z., Drozdzal, M., Romero, A., Rabbat,
  M., Vincent, P., Yakubova, N., Pinkerton, J., Wang, D., Owens, E., Zitnick,
  C.L., Recht, M.P., Sodickson, D.K., Lui, Y.W.: {fastMRI}: An open dataset and
  benchmarks for accelerated {MRI}. ArXiv e-prints  (2018)

\end{thebibliography}

\newpage

\section*{Details about Architectures}
\begin{figure}
\centering
\includegraphics[width=1.\textwidth]{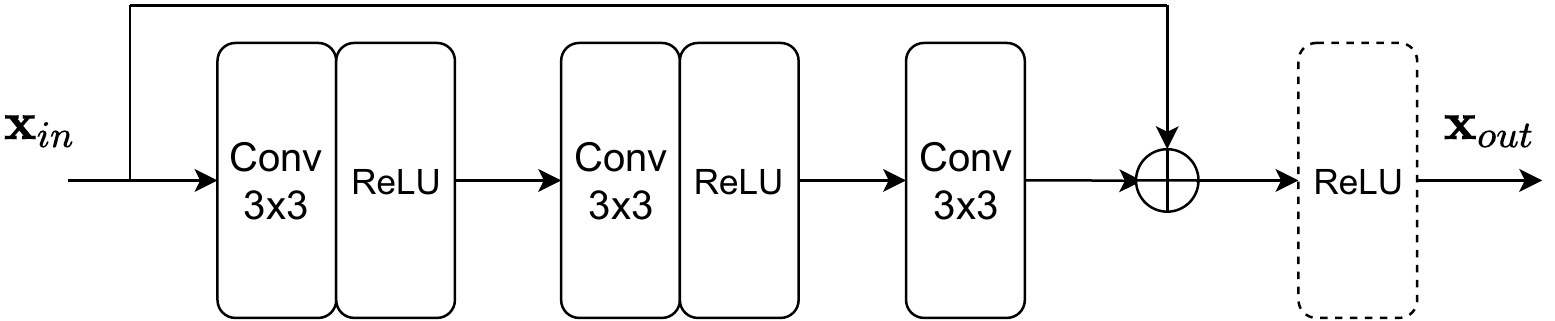}
\caption{Block diagram of a core residual block used in the proposed method and MoDL. Multiple blocks are serially concatenated to form the denoising networks. The final ReLU is replaced by a linear activation in the last block.}
\label{fig:core_resblock}
\end{figure}
Our model uses deep residual networks for both $\mathcal{D}_\mathbf{m}$ and $\mathcal{D}_\mathbf{s}$. The architecture of a residual block is shown in Fig. \ref{fig:core_resblock}, and multiple blocks are serially stacked to form the refinement networks. We use four blocks for both networks, leading to $447,000$ trainable parameters. For MoDL, we use a residual network identical to $\mathcal{D}_\mathbf{m}$ as an image denoiser, with a total number of $224000$ parameters. The same network is re-used across all outer unrolls. We use the open-source implementation of E2E-VarNet, with a number of $6$ cascades to match the number of outer unrolls for other methods, U-Nets with a depth of $3$ pooling layers, and a number of hidden channels of $12$ and $8$ for the image refinement and map estimation modules, respectively.

We treat complex-valued tensors as real-valued with two channels. All kernel sizes in all convolutional layers are $3\times3$ and the ReLU activation is used. Since Deep J-Sense assumes linear, not circular, convolution, we set the image size to include a region of padding. We use this padding to implement the k-space convolution in the image domain, by first converting both kernels (with padding) to the image domain, followed by a point-wise multiplication, conversion back to k-space and cropping. This achieves a linear convolution in k-space and avoids memory issues in PyTorch when convolutions with very large kernels are used. All models are trained using the Adam optimizer with default PyTorch parameters, learning rate $0.0002$, and batch size of $1$, for a total number of $30$ epochs. Every ten epochs, the learning rate is reduced by a factor of two. Additionally, we clip the gradients at each step to a maximum absolute value of $0.1$. We found that this was sufficient to achieve stable convergence for all the considered methods. Training our method for $30$ epochs takes approximately $20$ hours, while MoDL and E2E-VarNet take approximately $14$ hours. Training is carried out on a machine with two NVIDIA V100 GPUs.

We investigate the choice of the hyper-parameters $N$, $n_1$ and $n_2$ and how their trade-off impacts performance and find no significant difference in average or median performance between the configurations $(N, n_1, n_2) = $ $\{(3, 12, 12),$ $(4, 9, 9),$ $(6, 6, 6),$ $(9, 4, 4)\}$, but we do find a slight decrease if $n_1$ and $n_2$ are less than or equal to three. The comparisons in the main body are evaluated on a network with $N=n_1=n_2=6$. We choose a map kernel with dimensions $15\times9$, with the same aspect ratio as the field of view in fastMRI. We find that choosing a map kernel size larger than $25$ in either direction leads to instabilities in the end-to-end training.

\section*{Example Reconstructions}
\begin{figure}
\centering
\includegraphics[width=0.95\textwidth]{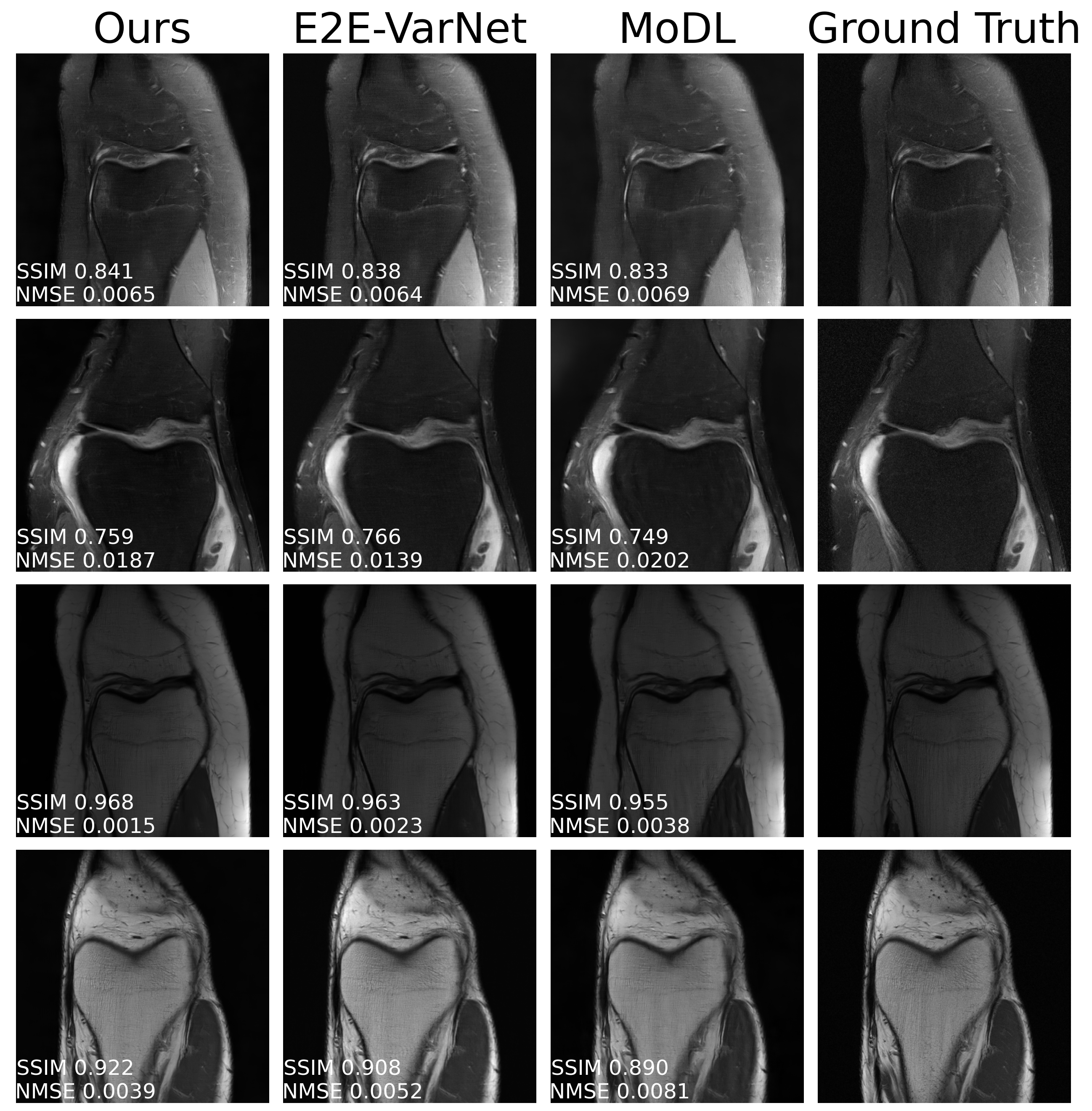}
\caption{Additional reconstruction examples at $R=4$. All images represent RSS images, central crop of $320\times320$.}
\label{fig:supp_scans}
\end{figure}

Fig. \ref{fig:supp_scans} shows multiple reconstruction results randomly chosen from the validation set. All cases show a significant qualitative and quantitative improvement against MoDL. The second row shows a sample where the proposed method is outperformed by E2E-VarNet.

\end{document}